\documentclass [10pt]{article}
\usepackage[left=2cm,top=2.50cm,right=2cm,bottom=2.50cm]{geometry}
\usepackage{mathrsfs}
\usepackage{amsmath,amssymb,latexsym,color,cancel,graphicx,bbm,colortbl}
\usepackage[english]{babel}
\usepackage[latin1]{inputenc}
\usepackage{ragged2e}
\usepackage{cite}

\begin{document}
\date{}

\title{Algebraic approach and Berry phase of a Hamiltonian with a general $SU(1,1)$ symmetry}
\author{E. Chore\~no$^{a}$, R. Valencia$^{a}$, D. Ojeda-Guill\'en$^{b}$\footnote{{\it E-mail address:} dojedag@ipn.mx}}
\maketitle

\begin{minipage}{0.9\textwidth}
\small $^{a}$ Escuela Superior de F{\'i}sica y Matem\'aticas,
Instituto Polit\'ecnico Nacional, Ed. 9, Unidad Profesional Adolfo L\'opez Mateos, Delegaci\'on Gustavo A. Madero, C.P. 07738, Ciudad de M\'exico, Mexico.\\

\small $^{b}$ Escuela Superior de C\'omputo, Instituto Polit\'ecnico Nacional,
Av. Juan de Dios B\'atiz esq. Av. Miguel Oth\'on de Mendiz\'abal, Col. Lindavista,
Delegaci\'on Gustavo A. Madero, C.P. 07738, Ciudad de M\'exico, Mexico.\\

\end{minipage}

\begin{abstract}
In this paper we study a general Hamiltonian with a linear structure given in terms of two different realizations of the $SU(1,1)$ group. We diagonalize this Hamiltonian by using the similarity transformations of the $SU(1,1)$ and $SU(2)$ displacement operators performed to the $su(1,1)$ Lie algebra generators. Then, we compute the Berry phase of a general time-dependent Hamiltonian with this general $SU(1,1)$ linear structure.\\[10pt]
\end{abstract}


\section{Introduction}

Group theory has become a very valuable tool when studying and solving various problems in theoretical physics. This theory has been applied in high-energy physics, condensed matter, atomic, molecular, and nuclear physics. Two of the main groups that are frequently used to describe these physical phenomena are the $SU(1,1)$ and $SU(2)$ groups. In particular, the $SU(1,1)$ and $SU(2)$ groups and their simple generalizations have been used to study many properties of relevant Hamiltonians in Quantum Optics like the Jaynes-Cummings model \cite{Jay}, the Tavis-Cummings model \cite{Dicke,TC} and the optical parametric amplifiers \cite{Louisell,Mollow}.

The Jaynes-Cummings model describes the interaction between radiation and matter, and is the simplest and completely soluble quantum-mechanical model. The exact solution of this theoretical model has been found in the rotating wave approximation \cite{Haroche}. Despite the simplicity of the Jaynes-Cummings model, it presents interesting quantum phenomena \cite{Narozhny,Kuklinski,Short,Diedrich,Milonni,Slosser,Gea,Phoenix}, all of them being experimentally corroborated, as can be seen in references \cite{Goy,Brune,Guerlin}.

The Tavis-Cummings model emerged from the study of $N$ identical two-level molecules interacting through a dipole coupling with a single-mode quantized radiation field at resonance. This model has been studied through different methods, among which are the Holstein-Primakoff transformation \cite{Bashir}, quantum inverse methods \cite{Bogoliubov,Rybin}, and polynomially deformed $su(2)$ algebras \cite{Vadeiko}. Both the Jaynes-Cummings model and the Tavis-Cummings model are still widely studied nowadays \cite{Lamata,Gerritsma,Lamata2,Retzker,Kopylov,Sun}.

In the optical parametric amplifier, one photon of a pump field transforms, via the nonlinear medium, into two photons called signal and idler. These output beams have the same frequency and polarization in the degenerate case and different ones in the non-degenerate case \cite{mandel}. In Ref. \cite{gerryberry}, Gerry used an $su(1,1)$ Lie algebra realization to study the Berry phase in the degenerate parametric amplifier.

The Berry phase \cite{Berry} is a phase factor gained by the wavefunction after the system is transported through a closed path via adiabatic variation of parameters. Since its introduction, it has been extensively studied in several quantum systems \cite{Xie,Bu,Thilagam}. We recently have applied the theory of the $SU(1,1)$ and $SU(2)$ groups to obtain the energy spectrum, eigenfunctions and the Berry phase of some of these Quantum Optics models \cite{Nos1,Nos2,Nos3,Nos4,Nos5}. In reference \cite{Nos6}, an algebraic method based on the $Sp(4,R)$ group (and contains the $SU(1,1)$ and $SU(2)$ groups) was introduced to solve exactly the interaction part of the most general Hamiltonian of a two-level system in two-dimensional geometry.

The aim of the present work is to introduce an algebraic method to solve exactly and compute the Berry phase of a general Hamiltonian with an $SU(1,1)$ symmetry.

This work is organized as follows. In Section 2, we construct a general Hamiltonian with a linear $SU(1,1)$ structure based on two different bosonic realizations of the $su(1,1)$ Lie algebra. Then, we introduce a method to diagonalize this Hamiltonian based on the similarity transformations of the $SU(1,1)$ and $SU(2)$ displacement operators performed to the $SU(1,1)$ generators. These transformations allow us to obtain the energy spectrum and eigenfunctions of this general Hamiltonian. In Section 3, we calculate the transformations of the operator $i\frac{\partial}{\partial t}$ in terms of the $SU(1,1)$ and $SU(2)$ displacement operators introduced in Section 2. With this result, we compute the Berry phase for a general time-dependent Hamiltonian with this $SU(1,1)$ linear structure. Section 4 is dedicated to study a particular case of the general $SU(1,1)$ Hamiltonian introduced in Section 2. Finally, we give some concluding remarks.

\section{A general linear $SU(1,1)$ bosonic Hamiltonian}

In many Quantum Optics problems related to parametric amplifiers we have the one-mode (also known as the ``squeezed oscillator Hamiltonian") and two-mode Hamiltonians
\begin{equation}
H_a=\omega \hat{a}^{\dag}\hat{a}+g\hat{a}^{\dag^2}+g^{*}\hat{a}^2,\label{Ha}
\end{equation}
\begin{equation}
H_{ab}=\omega_1 \hat{a}^{\dag}\hat{a}+\omega_2 \hat{b}^{\dag}\hat{b}+\lambda \hat{a}\hat{b}+\lambda^{*}\hat{a}^{\dag}\hat{b}^{\dag},\label{Hab}
\end{equation}
which are expressed in terms of the bosonic annihilation $\hat{a}$, $\hat{b}$ and creation $\hat{a}^{\dag}$, $\hat{b}^{\dag}$ operators. These operators obey the commutation relations
\begin{equation}
[\hat{a},\hat{a}^{\dag}]=[\hat{b},\hat{b}^{\dag}]=1,
\end{equation}
\begin{equation}
[\hat{a},\hat{b}]=[\hat{a}^{\dag},\hat{b}^{\dag}]=[\hat{a}^{\dag},\hat{b}]=[\hat{a},\hat{b}^{\dag}]=0.\label{boson}
\end{equation}
The Hamiltonians of equations (\ref{Ha}) and (\ref{Hab}) can be studied separately in terms of an appropriate $SU(1,1)$ realization. As it is well known, the $su(1,1)$ Lie algebra is defined in terms of the commutation relations \cite{Vourdas}
\begin{eqnarray}
[K_{0},K_{\pm}]=\pm K_{\pm},\quad\quad [K_{-},K_{+}]=2K_{0}.\label{com}
\end{eqnarray}
With the operators $\hat{a}^{\dag}\hat{a}$, $\hat{b}^{\dag}\hat{b}$, $\hat{a}^{\dag}\hat{b}^{\dag}$, $\hat{b}\hat{a}$, $\hat{a}^{\dag}{}^{2}$ and $\hat{a}^{2}$ we can construct the following two realizations of the $su(1,1)$ Lie algebra
\begin{equation}
K_{+}^{(ab)}=\hat{a}^{\dag}\hat{b}^{\dag},\quad\quad K_{-}^{(ab)}=\hat{b}\hat{a},\quad\quad K_{0}^{(ab)}=\frac{1}{2}(\hat{a}^{\dag}\hat{a}+\hat{b}^{\dag}\hat{b}+1)\quad\quad N_d^{(ab)}=\hat{a}^{\dag}\hat{a}-\hat{b}^{\dag}\hat{b}, \label{su11ab}
\end{equation}
and
\begin{equation}
K_{+}^{(a)}=\frac{1}{2}\hat{a}^{\dag}{}^{2},\quad\quad K_{-}^{(a)}=\frac{1}{2}\hat{a}^{2},\quad\quad K_{0}^{(a)}=\frac{1}{2}\left(\hat{a}^{\dag}\hat{a}+\frac{1}{2}\right). \label{su11a}
\end{equation}
Here, the operator $N_d^{(ab)}$ is the difference of the number operators of the two oscillators and commutes with all the generators of the algebra. Therefore, the $SU(1,1)$ Hamiltonians $H_a$ and $H_{ab}$ of equations (\ref{Ha}) and (\ref{Hab}) can be written as
\begin{equation}
H_a=\omega \left(2K_{0}^{(a)}-\frac{1}{2}\right)+2gK_{+}^{(a)}+2g^{*}K_{-}^{(a)},\label{Hak}
\end{equation}
\begin{equation}
H_{ab}=(\omega_1+\omega_2)\left(K_{0}^{(ab)}-\frac{1}{2}\right)+\frac{1}{2}(\omega_1-\omega_2)N_d^{(ab)}+\lambda K_{-}^{(ab)}+\lambda^{*}K_{+}^{(ab)}.\label{Habk}
\end{equation}

Similarly, the $su(2)$ Lie algebra is spanned by the generators $J_{+}$, $J_{-}$ and $J_{0}$, which satisfy the commutation relations \cite{Vourdas}
\begin{eqnarray}
[J_{0},J_{\pm}]=\pm J_{\pm},\quad\quad [J_{+},J_{-}]=2J_{0}.\label{com2}
\end{eqnarray}
With the bilinear products $\hat{a}^{\dag}\hat{a}$, $\hat{b}^{\dag}\hat{b}$, $\hat{a}^{\dag}\hat{b}$ and $\hat{b}^{\dag}\hat{a}$ we can construct an $su(2)$ Lie algebra realization by introducing the operators
\begin{equation}
J_+=\hat{a}^{\dag}\hat{b},\quad\quad J_-=\hat{b}^{\dag}\hat{a},\quad\quad J_0=\frac{1}{2}(\hat{a}^{\dag}\hat{a}-\hat{b}^{\dag}\hat{b}).\label{su2}
\end{equation}
Therefore, the operator $N_d^{(ab)}$ of the $su(1,1)$ Lie algebra is related to the $su(2)$ Lie algebra, since $N_d^{(ab)}=J_0$.

Based on all these results, we can introduce a more general Hamiltonian with an $SU(1,1 )$ linear structure
\begin{equation}
H=\omega \left(\hat{a}^{\dag}\hat{a}+\hat{b}^{\dag}\hat{b}\right)+\lambda \hat{a}\hat{b}+\lambda^{*}\hat{a}^{\dag}\hat{b}^{\dag}+g\hat{a}^{\dag^2}+g^{*}\hat{a}^2+c\hat{b}^{\dag^2}+c^{*}\hat{b}^2.
\end{equation}
We can write this general Hamiltonian in terms of the $su(1,1)$ Lie algebra realizations of equations (\ref{su11ab}) and (\ref{su11a}) as follows
\begin{align}
H&=\alpha_{-}^{(a)}K_{-}^{(a)}+\alpha_{+}^{(a)}K_{+}^{(a)}+\alpha_{-}^{(b)}K_{-}^{(b)}+\alpha_{+}^{(b)}K_{+}^{(b)}+\alpha_{-}^{(ab)}K_{-}^{(ab)}+\alpha_{+}^{(ab)}K_{+}^{(ab)}+\alpha_{0}^{(ab)}K_{0}^{(ab)},\label{GHsu11}
\end{align}
where the $\alpha$'s are complex constants such that $\alpha_{-}=\alpha^{*}_{+}$. This Hamiltonian can be diagonalized by using of the $SU(2)$ displacement operator $D(\chi)$ and the $SU(1,1)$ displacement operator $D(\xi)_{ab}=D(\xi_{a})D(\xi_{b})$ (see Appendix A). Here, the complex constants $\chi$, $\xi_a$ and $\xi_b$ are explicitly given by
\begin{equation}
\chi_=-\frac{\theta}{2}e^{-i\phi}; \quad \xi_{a}=-\frac{\theta_{a}}{2}e^{-i\phi_{a}}; \quad \xi_{b}=-\frac{\theta_{b}}{2}e^{-i\phi_{b}}.\label{complex}
\end{equation}
We use the $SU(2)$ and $SU(1,1)$ displacement operators to transform each of the operators of the different $su(1,1)$ bosonic realizations $K_{\pm}^{(a)}$, $K_{\pm}^{(b)}$, $K_{\pm}^{(ab)}$ and $K_{0}^{(ab)}$, as it is shown in the Appendix A. Thus, in order to remove the ladder operators $K^{(ab)}_{+}$ and $K^{(ab)}_{-}$ of the Hamiltonian (\ref{GHsu11}), we first apply the similarity transformation in terms of $D(\chi)$ as
\begin{equation}
H'=D^{\dag}(\chi)HD(\chi). \label{Hprima}
\end{equation}
By using the equations (\ref{transf1})-(\ref{transf4}) we can write the new Hamiltonian $H'$ as
\begin{equation}
H'=\beta_{-}^{(a)}K_{-}^{(a)}+\beta_{+}^{(a)}K_{+}^{(a)}+\beta_{-}^{(b)}K_{-}^{(b)}+\beta_{+}^{(b)}K_{+}^{(b)}+\beta_{-}^{(ab)}K_{-}^{(ab)}+\beta_{+}^{(ab)}K_{+}^{(ab)}+\beta_{0}^{(ab)}K_{0}^{(ab)},\label{H'}
\end{equation}
where $(\beta_{\pm}^{(i)})^{\dag}=\beta_{\mp}^{(i)}$ and the new complex constants $\beta$'s are given as
\begin{align}&
\beta_{-}^{(a)}=\frac{1}{2}\alpha_{-}^{(a)}(\cos{(2|\chi|)}+1)-\frac{(\chi^{*})^{2}}{2|\chi|^{2}}\alpha_{-}^{(b)}(\cos{(2|\chi|)}-1)-\alpha_{-}^{(ab)}\frac{\chi^{*}}{|\chi|}\sin{(2|\chi|)}, \label{betaA}\\&
\beta_{-}^{(b)}=-\frac{\chi^{2}}{2|\chi|^{2}}\alpha_{-}^{(a)}(\cos{(2|\chi|)}-1)+\frac{1}{2}\alpha_{-}^{(b)}(\cos{(2|\chi|)}+1)+\alpha_{-}^{(ab)}\frac{\chi}{|\chi|}\sin{(2|\chi|)},\label{betaB}\\&
\beta_{-}^{(ab)}= \frac{\chi}{2|\chi|}\alpha_{-}^{(a)}\sin{(2|\chi|)}-\frac{\chi^{*}}{2|\chi|}\alpha_{-}^{(b)}\sin{(2|\chi|)}+\alpha_{-}^{(ab)}\cos{(2|\chi|)},\label{betaAB}\\&
\beta_{0}^{(ab)}=\alpha_{0}^{(ab)}=\alpha_{0}. \label{beta0}
\end{align}
Therefore, if we choose the parameters of the complex number $\chi=-\frac{\theta}{2}e^{-i\phi}$ as
\begin{eqnarray}
\frac{\tan{\theta}}{2}=\frac{\sqrt{(\alpha_{+}^{(ab)}\alpha_{-}^{(a)}+\alpha_{-}^{(ab)}\alpha_{+}^{(b)})(\alpha_{+}^{(ab)}\alpha_{-}^{(b)}+\alpha_{-}^{(ab)}\alpha_{+}^{(a)})}}{\alpha_{-}^{(a)}\alpha_{+}^{(a)}-\alpha_{-}^{(b)}\alpha_{+}^{(b)}}, \quad
e^{i\phi}=\sqrt{\frac{\alpha_{+}^{(ab)}\alpha_{-}^{(a)}+\alpha_{-}^{(ab)}\alpha_{+}^{(b)}} {\alpha_{+}^{(ab)}\alpha_{-}^{(b)}+\alpha_{-}^{(ab)}\alpha_{+}^{(a)}}},\label{Cond1}
\end{eqnarray}
we can eliminate the coefficients $\beta_{\pm}^{(ab)}$, and the Hamiltonian of the equation (\ref{H'}) is reduced to
\begin{equation}
H'=\beta_{0}^{(ab)}K_{0}^{(ab)}+\beta_{-}^{(a)}K_{-}^{(a)}+\beta_{+}^{(a)}K_{+}^{(a)}+\beta_{-}^{(b)}K_{-}^{(b)}+\beta_{+}^{(b)}K_{+}^{(b)}.
\end{equation}

Following the above procedure, we now apply the similarity transformation in terms of the displacement operator $D(\xi)_{ab}$ to this Hamiltonian $H'$ as follows
\begin{equation}
H''=D^{\dag}(\xi)_{ab}H'D(\xi)_{ab}. \label{Hbiprima}
\end{equation}
Given that the boson operators $\hat{a}$ and $\hat{b}$ commute and by using the equations (\ref{transf5}) and (\ref{transf6}) of the Appendix A, we can show that the Hamiltonian $H''$ is transformed to
\begin{equation}
H''=\sqrt{\alpha_{0}^{2}-4\beta_{+}^{(a)}\beta_{-}^{(a)}}K_{0}^{(a)}+\sqrt{\alpha_{0}^{2}-4\beta_{+}^{(b)}\beta_{-}^{(b)}}K_{0}^{(b)}.\label{H''}
\end{equation}
Here, the parameters of the complex numbers $\xi_{a}=-\frac{\theta_{a}}{2}e^{-i\phi_{a}}$ and $\quad \xi_{b}=-\frac{\theta_{b}}{2}e^{-i\phi_{b}}$ were chosen as
\begin{eqnarray}
\tanh{(\theta_{i})}=\frac{2}{\alpha_{0}}\sqrt{\beta_{-}^{(i)}\beta_{+}^{(i)}}; \quad \quad e^{i\phi_{i}}=\sqrt{\frac{\beta_{-}^{(i)}}{\beta_{+}^{(i)}}},\label{Cond2}
\end{eqnarray}
with $i=a,b$.

From expression (\ref{H''}) we can see that the eigenstates $|\varphi''\rangle$ of the Hamiltonian $H''$ are the direct product of the number states of the modes $\{a,b\}$, that is $|\varphi''\rangle=|n_{a}\rangle\otimes|n_{b}\rangle$. Hence, these states explicitly are the eigenfunctions of the two-dimensional harmonic oscillator
\begin{equation}
\varphi''_{n_{l},m_n}(\rho,\phi)=\frac{1}{\sqrt{\pi}}e^{im_n\phi}(-1)^{n_{l}}\sqrt{\frac{2(n_{l})!}{(n_{l}+m_n)!}}\rho^{m_n}L^{m_n}_{n_{l}}(\rho^{2})e^{-1/2\rho^{2}}\label{Poly1},
\end{equation}
where $n_l$ is the left chiral quantum number. From this result we obtain that the energy spectrum of the Hamiltonian $H''$, and therefore of the general $SU(1,1)$ Hamiltonian $H$ of equation (\ref{GHsu11}), is given by
\begin{equation}
E_{n_{l},m_{n}}=\left(\sqrt{\alpha_{0}^{2}-4\beta_{+}^{(a)}\beta_{-}^{(a)}}+\sqrt{\alpha_{0}^{2}-4\beta_{+}^{(b)}\beta_{-}^{(b)}}\right)\frac{(n_{l}+1)}{4}+\left(\sqrt{\alpha_{0}^{2}-4\beta_{+}^{(a)}\beta_{-}^{(a)}}-\sqrt{\alpha_{0}^{2}-4\beta_{+}^{(b)}\beta_{-}^{(b)}}\right)\frac{m_{n}}{4}\label{E}.
\end{equation}
Moreover, the eigenfunctions of this general $SU(1,1)$ Hamiltonian $H$ are obtained from the relationship
\begin{equation}
|\varphi\rangle=D(\chi)D(\xi)_{ab}|\varphi''\rangle,\label{estado}
\end{equation}
where the term $D(\xi)_{ab}|\varphi''\rangle$ can be identified as the $SU(1,1)$ Perelomov number coherent states for the two-dimensional harmonic oscillator \cite{Nos5}.

\section{The Berry phase of the general $SU(1,1)$ time-dependent Hamiltonian}

In this Section we shall now consider the Hamiltonian (\ref{GHsu11}) as an explicit function of time $H(t)$, that is
\begin{equation}
H(t)=\alpha_{0}(t)K_{0}^{(ab)}+\alpha_{+}^{(ab)}(t)K_{+}^{(ab)}+{\alpha_{-}^{(ab)}(t)}K_{-}^{(ab)}+\alpha_{+}^{(a)}(t)K_{+}^{(a)}+{\alpha_{-}^{(a)}(t)}K_{-}^{(a)}+\alpha_{+}^{(b)}(t)K_{+}^{(b)}+{\alpha_{-}^{(b)}(t)}K_{-}^{(b)}\label{Ht},
\end{equation}
where the $\alpha$'s are complex constants such that $\alpha^{(j)}_{\pm}=\left(\alpha^{(j)}_{\mp}\right)^{*}$ and can be written as
\begin{equation}
\alpha_{+}^{(j)}(t)=\lambda_{j}(t)e^{i\gamma_{j}(t)}.
\end{equation}
Here $\lambda_{j}^{(1)}(t)$ and $\phi_{j}^{(1)}(t)$ with $j=a,b,ab,$ are arbitrary real functions of time. Since this Hamiltonian is time-dependent, to describe quantum dynamics we shall use the Schr\"odinger picture
\begin{equation}
i\hbar\frac{d}{dt}|\psi(t)\rangle=H(t)|\psi(t)\rangle.\label{Schr}
\end{equation}
Thus, in order to study the time evolution of the states of Hamiltonian (\ref{Ht}), we will use the time-dependent nontrivial invariant Hermitian operator $I(t)$ \cite{Lewis1,Lewis2}, which satisfies the conditions
\begin{equation}
i\frac{\partial}{\partial t}I(t)+[I(t),H(t)]=0.\label{inavariante}
\end{equation}
Now, we shall use the time-dependent versions of the $SU(1,1)$ and $SU(2)$ displacement operators of equations (\ref{Dsu11xsu11}) and (\ref{Dsu2}), where the complex parameters $\theta$'s and $\phi$'s of the expressions (\ref{complex}) are arbitrary real functions of time. Then, with these considerations we can define the invariant operator $I(t)$ as
\begin{equation}
I(t)=D(\chi(t))D(\xi(t))_{ab}K_{0}^{(ab)}D^{\dag}(\xi(t))_{ab}D^{\dag}(\chi(t)),\label{opinva}
\end{equation}
or explicitly as
\begin{equation}
I(t)=\beta_{0}J_{0}+\beta_{1}J_{+}+\beta_{1}^{*}J_{-}+\beta_{2}K_{0}^{(ab)}+\beta_{3}K_{+}^{(ab)}+\beta_{3}^{*}K_{-}^{(ab)}+\beta_{4}K_{+}^{(a)}+\beta_{4}^{*}K_{-}^{(a)}+\beta_{5}K_{+}^{(b)}+\beta_{5}^{*}K_{+}^{(b)},\label{I}
\end{equation}
where the $\beta$'s coefficients are given by
\begin{align}
&\beta_{0}=\frac{\cosh(\theta_{a})-\cosh(\theta_{2})}{2}\cos(\theta),\\&
\beta_{1}=\frac{\cosh(\theta_{a})-\cosh(\theta_{2})}{4}\sin(\theta)e^{i\phi},\\&
\beta_{2}=\frac{\cosh(\theta_{a})+\cosh(\theta_{2})}{2},\\&
\beta_{3}=\sin(\theta)\left(\sinh(\theta_{a})e^{i(\phi-\phi_{a})}-\sinh(\theta_{b})e^{-i(\phi+\phi_{b})}\right),\\&
\beta_{4}=\sinh(\theta_{a})\left(\cos(\theta)+1\right)e^{-i\phi_{a}}-\sinh(\theta_{b})\left(\cos(\theta)-1\right)e^{-2i\phi},\\&
\beta_{5}=\sinh(\theta_{b})\left(\cos(\theta)+1\right)e^{-i\phi_{b}}-\sinh(\theta_{a})\left(\cos(\theta)-1\right)e^{2i\phi}.
\end{align}
From the condition of equation (\ref{inavariante}) and the form of the invariant operator $I(t)$, we obtain that the time-dependent physical parameters of the complex constants $\alpha^{j}_{\pm}$ are related to coefficients $\beta$'s as follows
\begin{align}
&\dot{\beta_{0}}+\alpha^{(a)}_{+}\beta^{*}_{4}-\alpha^{(a)}_{-}\beta_{4}-\alpha^{(b)}_{+}\beta^{*}_{5}+\alpha^{(b)}_{-}\beta_{5}=0,\nonumber\\ &
\dot{\beta_{1}}+\alpha^{(ab)}_{+}\beta^{*}_{5}+\alpha^{(a)}_{+}\beta^{*}_{3}-\alpha^{(ab)}_{-}\beta_{4}-\alpha^{(b)}_{-}\beta_{3}=0,\nonumber\\&
\dot{\beta_{2}}+2\alpha^{(ab)}_{+}\beta^{*}_{3}-2\alpha^{(ab)}_{-}\beta_{3}+\alpha^{(a)}_{+}\beta^{*}_{4}-\alpha^{(a)}_{-}\beta_{4}+\alpha^{(b)}_{+}\beta^{*}_{5}-\alpha^{(b)}_{-}\beta_{5}=0,\nonumber\\&
\dot{\beta_{3}}+\alpha^{(b)}_{+}\beta_{1}+\alpha^{(a)}_{+}\beta^{*}_{1}+\alpha^{(ab)}_{+}\beta_{2}-\alpha_{0}\beta_{3}=0,\nonumber\\&
\dot{\beta_{4}}+\alpha^{(a)}_{+}\beta_{0}+2\alpha^{(ab)}_{+}\beta_{1}+\alpha^{(a)}_{+}\beta_{2}-\alpha_{0}\beta_{4}=0,\nonumber\\&
\dot{\beta_{5}}+\alpha^{(b)}_{+}\beta_{2}+2\alpha^{(ab)}_{+}\beta^{*}_{1}-\alpha^{(b)}_{+}\beta_{0}-\alpha_{0}\beta_{5}=0,\label{dotequations}
\end{align}
together with their corresponding conjugate equations.

On the other hand, the transformations of the algebra generators $K$'s and $J$'s under time-dependent displacement operators $D(\xi(t))_{ab}$ and $D(\chi(t))$ remain unchanged and are given by the expressions of the Appendix A. In addition, we can transform the operator $i\frac{\partial}{\partial t}$ under the time-dependent displacement operators $D(\xi(t))_{ab}$ and $D(\chi(t))$ as
\begin{equation}
D^{\dag}(\xi(t))_{ab}D^{\dag}(\chi(t))\left(i\frac{\partial}{\partial t}\right)D(\chi(t))D(\xi(t))_{ab}=\left(i\frac{\partial}{\partial t}\right)''.
\end{equation}
to obtain
\begin{align}
\left(i\frac{\partial}{\partial t}\right)''&=i\frac{\partial}{\partial t}+ (b_{0}+c_{0})K_{0}^{(ab)}+(b_{1}+c_{2})J_{0}+c_{3}J_{+}+c_{3}^{*}J_{-}+c_{4}K_{+}^{(ab)}+c_{4}^{*}K_{-}^{(ab)}\nonumber\\&+(b_{2}+c_{5})K_{+}^{(a)}+(b_{2}^{*}+c_{5}^{*})K_{-}^{(a)}+(b_{3}+c_{6})K_{+}^{(b)}+(b_{3}^{*}+c_{6}^{*})K_{-}^{(b)}.\label{dt}
\end{align}
The explicit form of the constants $b$'s and $c$'s are given in the Appendix B. Also, as it is shown in Ref. \cite{Lewis2}, if the eigenstates of the invariant operator satisfy the Schr\"odinger equation its eigenvalues are real. Therefore, given that $K_{0}|k,n\rangle=(k+n)|k,n\rangle$ we have
\begin{equation*}
D(\chi)D(\xi)_{ab}K^{(ab)}_{0}|k,n\rangle=(k+n)D(\chi)D(\xi)_{ab}|k,n\rangle,
\end{equation*}
which implies that
\begin{equation*}
 I(t)D(\chi)D(\xi)_{ab}|k,n\rangle=(k+n)D(\chi)D(\xi)_{ab}|k,n\rangle.
\end{equation*}
Thus, the states of the invariant operator $I(t)$ are $D(\chi)D(\xi)_{ab}|k,n\rangle=D(\chi)|\zeta(t),k,n\rangle$ where $|\zeta(t),k,n\rangle$ are the $SU(1,1)$ Perelomov number coherent states.

Moreover, if the states $|\psi(t)\rangle$ satisfy the relationship (\ref{Schr}) for the Hamiltonians $H(t)$, these states can be expanded through the sates $D(\chi)|\zeta(t),k,n\rangle$ in the form
\begin{equation}
|\psi(t)\rangle_{su(1,1)}=\sum_{n}a_{n}e^{i\alpha_{n}}D(\chi)|\zeta(t),k,n\rangle,
\end{equation}
where according to Lewis \cite{Lewis2} the phase $\alpha$ is given as
\begin{equation}
\alpha=\int_{0}^{t}dt'\langle \lambda,\kappa|i\frac{\partial}{\partial t'}-H(t')|\lambda,\kappa\rangle.\label{T-phase}
\end{equation}
Here, $|\lambda,\kappa\rangle$ are the eigenstates and $\lambda$ are the eigenvalues of the invariant operator $I(t)$.
Therefore, the phase of the eigenstates $D(\chi)|\zeta(t),k,n\rangle$ in a non-adiabatic process is given by
\begin{equation}
\alpha_{n,\mu}=(n+k)\int_{0}^{t}\left[b_{0}(t)+c_{0}(t)-\frac{A_{0}(t)+B_{0}(t)}{2}\right]dt'-\mu\int_{0}^{t}\left[\frac{A_{0}(t)+B_{0}(t)}{2}\right]dt',\label{nophase1}
\end{equation}
where as it is shown in the Ref. \cite{Berry}, the terms $A_{0}(t)$ and $B_{0}(t)$ are given as
\begin{align}
&A_{0}(t)=\cosh(\theta_{a}(t))\alpha_{0}-2\lambda'_{a}\sinh(\theta_{a}(t))\cos(\phi_{a}(t)+\gamma'_{a}(t)),\\ \nonumber & B_{0}(t)=\cosh(\theta_{b}(t))\alpha_{0}-2\lambda'_{b}\sinh(\theta_{b}(t))\cos(\phi_{b}(t)+\gamma'_{b}(t)).
\end{align}
It is worth mentioning that  for simplicity we have taken the $\beta$'s coefficients of the Hamiltonian (\ref{H''}) as
\begin{equation}
\beta_{+}^{(j)}(t)=\lambda'_{j}(t)e^{i\gamma'_{j}(t)},
\end{equation}
where $\lambda'_{j}(t)$ and $\phi'_{j}(t)$ with $j=a,b,$ are arbitrary real functions of time.

Unlike in a non-adiabatic process, in an adiabatic process we have that $\dot{\theta}=\dot{\phi}=0$. Therefore, from the relations (\ref{dotequations}) we can obtain the time-dependent versions of the expressions (\ref{Cond1}) and (\ref{Cond2}). Therefore, in an adiabatic process the phase of the states $D(\chi)|\zeta(t),k,n\rangle$ are reduced to
\begin{align}
\alpha_{n_{l},m_{n}}&=-\frac{(n_{l}+1)}{4}\int_{0}^{t}\left(\sqrt{\alpha_{0}(t')^{2}-4\beta_{+}^{(a)}(t')\beta_{-}^{(a)}(t')}+\sqrt{\alpha_{0}(t')^{2}-4\beta_{+}^{(b)}(t')\beta_{-}^{(b)}(t')}\right)dt'\nonumber\\ &-\frac{m_{n}}{4}\int_{0}^{t}\left(\sqrt{\alpha_{0}(t')^{2}-4\beta_{+}^{(a)}(t')\beta_{-}^{(a)}(t')}-\sqrt{\alpha_{0}(t')^{2}-4\beta_{+}^{(b)}(t')\beta_{-}^{(b)}(t')}\right)dt'.\label{dy}
\end{align}
These are known as the dynamical phases and are defined as
\begin{equation}
\dot{\epsilon}_{n}=\langle\lambda,\kappa|H(t')|\lambda,\kappa\rangle\label{dyphase}.
\end{equation}
while the Berry phase is defined as
\begin{equation}
\dot{\gamma}_{\kappa}=i\langle\lambda,\kappa|\frac{\partial}{\partial t}|\lambda,\kappa\rangle.\label{berphase}
\end{equation}
Thus, the Berry phase of the states $D(\chi)|\zeta(t),k,n\rangle$ is obtained in the adiabatic limit as follows
\begin{equation}
\gamma_{n,\mu}(T)=(n+k)\int_{0}^{T}(b_{0}+c_{0})dt+\mu\int_{0}^{T}(b_{1}+c_{1})dt,\label{Berrysu11}
\end{equation}
where $T$ denotes the period. According to the values of the constants $b$ and $c$ shown in the Appendix B, the Berry phase is
\begin{align}
\gamma_{k,n,\mu}(T)&=\frac{(n+k)}{2}\left[\int_{0}^{T}\dot{\phi}_{a}(\cosh(\theta_{a})-1)dt+\int_{0}^{T}\dot{\phi}_{b}(\cosh(\theta_{b})-1)dt+\int_{0}^{T}\dot{\phi}\left(\cos(\theta)-1\right)\left(\cosh(\theta_{a})-\cosh(\theta_{b})\right)dt\right]\nonumber\\&+\frac{\mu}{2}\left[\int_{0}^{T}\dot{\phi}_{a}(\cosh(\theta_{a})-1)dt-\int_{0}^{T}\dot{\phi}_{b}(\cosh(\theta_{b})-1)dt+\int_{0}^{T}\dot{\phi}\left(\cos(\theta)-1\right)\left(\cosh(\theta_{a})+\cosh(\theta_{b})\right)dt\right].
\end{align}

To see the topological aspect of the Berry phase explicitly for our problem, let us suppose that the $\gamma_{ab}$, $\gamma_{a}$ and $\gamma_{b}$ phases are not independent of each other but they are related as follows $2\gamma_{ab}-\gamma_{a}-\gamma_{b}=n\pi$. If $n=0$, the coherent parameters of the  $\xi_{b}$, $\xi_{a}$ and $\chi$ complex constants are related to $\lambda'_{j}$ and $\gamma'_{j}$ physical constants of the $\alpha^{j}_{\pm}$ coefficients as
\begin{equation}
\phi=\gamma_{b}-\gamma_{ab}=\gamma_{ab}-\gamma_{a},\quad\quad\phi_{a}=-\gamma_{a},\quad\quad\phi_{b}=-\gamma_{b},
\end{equation}
and
\begin{equation}
\theta=\tan^{-1}\left[\frac{2\lambda_{ab}}{\lambda_{a}-\lambda_{b}}\right],\quad\quad\theta_{a}=f_{a}\left(\lambda_{ab},\lambda_{a},\lambda_{b}\right),\quad\quad\theta_{b}=f_{b}\left(\lambda_{ab},\lambda_{a},\lambda_{b}\right).
\end{equation}
Hence, the Berry phase is reduced to the closed integral
\begin{align}
\gamma_{k,n,\mu}(C)&=\frac{(n+k)}{2}\left[(\cosh(\theta_{a})-1)\oint d\gamma_{a}+(\cosh(\theta_{b})-1)\oint d\gamma_{b}+\left(\cos(\theta)-1\right)\left(\cosh(\theta_{a})-\cosh(\theta_{b})\right)\oint(d\gamma_{ab}-d\gamma_{a})\right]\nonumber\\&+\frac{\mu}{2}\left[(\cosh(\theta_{a})-1)\oint d\gamma_{a}-(\cosh(\theta_{b})-1)\oint d\gamma_{b}+\left(\cos(\theta)-1\right)\left(\cosh(\theta_{a})+\cosh(\theta_{b})\right)\oint(d\gamma_{ab}-d\gamma_{a})\right].
\end{align}
Therefore, the Berry phase of the states $D(\chi)|\zeta(t),k,n\rangle$ is finally given by
\begin{align}
\gamma_{k,n,\mu}(C)&=2\pi\frac{\mu}{2}\left[(\cosh(\theta_{b})-\cosh(\theta_{a})\right]-2\pi\frac{(n+k)}{2}\left[\cosh(\theta_{a})+\cosh(\theta_{b})-2\right]\label{Berry}.
\end{align}
It is obvious that in the cases where the condition $2\gamma_{ab}(t)-\gamma_{a}(t)-\gamma_{b}(t)=n\pi$ is satisfied, the Berry phases do not depend on an explicit form of the functions $\gamma_{ab}(t)$, $\gamma_{a}(t)$ and $\gamma_{b}(t)$, which are part of the complex coefficients of the Hamiltonian (\ref{Ht}).

\section{Generalized two-mode harmonic oscillator model}

The time-independent Hamiltonian that considers all possible linear interactions in the phase space of moments and positions is given by the expression \cite{Xie,Cervero}
\begin{equation}
H=H_{0}+\sum_{i=1,2}\left[\frac{\omega_{i}u_{i}}{2}(x_{i}p_{i}+p_{i}x_{i})\right]+s\frac{p_{1}p_{2}}{2m}+\sqrt{\omega_{1}\omega_{2}}\left(ux_{1}p_{2}+u'x_{2}p_{1}\right)+\frac{\omega_{1}\omega_{2}}{2}mvx_{1}x_{2},
\end{equation}
where $H_{0}$ is the Hamiltonian of the two-dimensional harmonic oscillator, $u_{i}$, $u$, $u'$, $s$, $v$ are real constants, and $\omega_{1}$, $\omega_2$ are the oscillation frequencies. By introducing the bosonic operators $a_{i}=(m\omega_{i}x_{i}+ip_{i})/\sqrt{2m\omega_{i}}$ and using the realizations (\ref{su11ab}), (\ref{su11a}) and (\ref{su2}), the above Hamiltonian can be rewritten as \cite{Nos6}
\begin{equation}
H=\beta_{0}J_{0}+\beta_{+}J_{+}+\beta_{-}J_{-}+\alpha_{0}K_{0}^{(12)}+\alpha_{+}^{(12)}K_{+}^{(12)}+{\alpha_{-}^{(12)}}K_{-}^{(12)}+\alpha_{+}^{(1)}K_{+}^{(1)}+{\alpha_{-}^{(1)}}K_{-}^{(1)}+\alpha_{+}^{(2)}K_{+}^{(2)}+{\alpha_{-}^{(2)}}K_{-}^{(2)},\label{CH}
\end{equation}
where the $\alpha's$ and $\beta's$ are complex constants such that $\alpha_{+}=\alpha^{*}_{-}$ and $\beta_{+}=\beta^{*}_{-}$. Here, these complex constants are given by the following expressions
\begin{equation}
\alpha_{0}=\omega_{1}+\omega_{2},\quad\quad\beta_{0}=\omega_{1}-\omega_{2},
\end{equation}
\begin{equation}
\alpha^{(1)}_{+}=-\frac{i\omega_{1}}{2}u_{1},\quad\quad\alpha^{(2)}_{+}=-\frac{i\omega_{2}}{2}u_{2},
\end{equation}
\begin{equation}
\alpha^{(12)}_{+}=\frac{\sqrt{\omega_{1}\omega_{2}}}{8}(v-s)-\frac{i}{4}(\omega_{2}u+\omega_{1}u'),\quad\quad\beta_{+}=\frac{\sqrt{\omega_{1}\omega_{2}}}{8}(v+s)+\frac{i}{4}(\omega_{2}u-\omega_{1}u').
\end{equation}

The general harmonic oscillator Hamiltonian of equation (\ref{CH}) can be mapped onto a Hamiltonian of the type (\ref{GHsu11}). Choosing the real constants $ u $, $ u '$, $ v $, $ s $ so that $ u = u' $, $ v = -s $, and taking the isotropic case ($\omega_{1}=\omega_{2}$), we have that the Hamiltonian (\ref{CH}) can be written as
\begin{equation}
H=\alpha_{0}K_{0}^{(12)}+\alpha_{+}^{(12)}K_{+}^{(12)}+{\alpha_{-}^{(12)}}K_{-}^{(12)}+\alpha_{+}^{(1)}K_{+}^{(1)}+{\alpha_{-}^{(1)}}K_{-}^{(1)}+\alpha_{+}^{(2)}K_{+}^{(2)}+{\alpha_{-}^{(2)}}K_{-}^{(2)},\label{CHSU11}
\end{equation}
where now the complex constants $\alpha's$ are given by
\begin{equation}
\alpha_{0}=2\omega,\quad\alpha^{(1)}_{+}=-i\frac{u_{1}\omega}{2},\quad\alpha^{(2)}_{+}=-i\frac{u_{2}\omega}{2},\quad\alpha^{(12)}_{+}=\frac{\omega_{1}}{4}v-i\frac{\omega}{2}u.
\end{equation}
Therefore, the Hamiltonian (\ref{CHSU11}) could be considered as a particular case of the Hamiltonian of the two-dimensional isotropic harmonic oscillator with linear interactions in the 4-dimension $x-p$ phase. On the other hand, if the constants $v$, $u$, $u_{1}$ and $u_{2}$ are arbitrary real functions of time which vary smoothly with time, its respective dynamical and Berry phases in the adiabatic limit are given by the expressions (\ref{dy}) and (\ref{Berry}), respectively.

\section{Concluding remarks}

In this paper we introduced a general Hamiltonian with a general linear structure given in terms of two different realizations of the $SU(1,1)$ group. We developed a method to diagonalize this Hamiltonian based on the similarity transformations of the $SU(1,1)$ and $SU(2)$ displacement operators performed to the $SU(1,1)$ generators. With these transformations, we were able to obtain the energy spectrum and eigenfunctions of our general $SU(1,1)$ Hamiltonian. Then, by using the similarity transformations of the operator $i\frac{\partial}{\partial t}$, we computed the Berry phase of a general time-dependent Hamiltonian with this $SU(1,1)$ linear structure.

It is important to note that, even though our general Hamiltonian was written only in terms of the $SU(1,1)$ group, it was necessary to introduce the $SU(2)$ group theory to be able to diagonalize it. This fact can be explained by remembering that the $SU(1,1)$ and $SU(2)$ groups, together with the so-called potential group $SU_p(1,1)$ can be imbedded into a larger group, $Sp(4,R)$. Also, since the Hamiltonian studied in this paper is very general, our results can be adequately transferred to more specific problems with these symmetries, such as the degenerate and non-degenerate parametric amplifier, among others.

\section*{Acknowledgments}

This work was partially supported by SNI-M\'exico, EDI-IPN, SIP-IPN Project Number $20200225$.

\section*{Appendix A. The similarity transformations of the $su(1,1)$ Lie algebra realizations}

In this Appendix we shall compute some similarity transformations of the operators $K_{\pm}^{(a)}$, $K_{\pm}^{(b)}$, $K_{\pm}^{(ab)}$ and $K_{0}^{(ab)}$ of the different $su(1,1)$ Lie algebra realizations introduced in Section 2. Therefore, by introducing the $SU(2)$ displacement operator
\begin{equation}
D(\chi)= \exp[\chi J_{+}-\chi^{*}J_{-}],\label{Dsu2}
\end{equation}
and considering the commutation relations of the $Sp(4,R)$ Lie algebra \cite{Nos6}, we can find the following results
\begin{equation}
D^{\dag}(\chi)K_{-}^{(a)}D(\chi)= \frac{\chi}{2|\chi|} K_{-}^{(ab)}\sin(2|\chi|) +\frac{1}{2}K_{-}^{(a)}\left(\cos(2|\chi|)+1\right)-\frac{\chi^{2}}{2|\chi|^{2}}K_{-}^{(b)}\left(\cos(2|\chi|)-1\right),\label{transf1}
\end{equation}
\begin{equation}
D^{\dag}(\chi)K_{-}^{(b)}D(\chi)=\frac{-\chi^{*}}{2|\chi|}K_{-}^{(ab)}\sin(2|\chi|) -\frac{(\chi^{*})^{2}}{2|\chi|^{2}} K_{-}^{(a)}\left(\cos(2|\chi|)-1\right) +\frac{1}{2}K_{-}^{(b)}\left(\cos(2|\chi|)+1\right), \label{transf2}
\end{equation}
\begin{equation}
D^{\dag}(\chi)K_{-}^{(ab)}D(\chi)= K_{-}^{(ab)}\cos(2|\chi|)-\frac{\chi^{*}}{|\chi|}K_{-}^{(a)}\sin(2|\chi|)+ \frac{\chi}{|\chi|} K_{-}^{(b)}\sin(2|\chi|),\label{transf3}
\end{equation}
\begin{equation}
D^{\dag}(\chi)K_{0}^{(ab)}D(\chi)=K_{0}^{(ab)} \label{transf4}.
\end{equation}
Similarly, we can introduce the $SU(1,1)\times SU(1,1)$ displacement operator as follows
\begin{equation}
D(\xi)_{ab}=D(\xi_{a})D(\xi_{b})= \exp[\xi_{a}K_{+}^{(a)}-\xi_{a}^{*}K_{-}^{(a)}+\xi_{b}K_{+}^{(b)}-\xi_{b}^{*}K_{-}^{(b)}].\label{Dsu11xsu11}
\end{equation}
Thus, the similarity transformations of the operators $K_{\pm}^{(a)}$, $K_{\pm}^{(b)}$, $K_{\pm}^{(ab)}$ and $K_{0}^{(ab)}$ in terms of this displacement operator are presented below
\begin{equation}
D^{\dag}(\xi)_{ab}K_{+}^{(i)}D(\xi)_{ab}=\sinh(2|\xi_{i}|)\frac{\xi_{i}^{*}}{|\xi_{i}|} K_{0}^{(i)}+\left(\cosh(2|\xi_{i}|)+1\right)\frac{K_{+}^{(i)}}{2}+\left(\cosh(2|\xi_{i}|)-1\right)\frac{\xi_{i}^{*}}{2\xi_{i}}K_{-}^{(i)},\label{transf5}
\end{equation}
\begin{equation}
D^{\dag}(\xi)_{ab}K_{0}^{(i)}D(\xi)_{ab}=\cosh(2|\xi_{i}|)K_{0}^{(i)}+\sinh(2|\xi_{i}|)\left(\frac{\xi_{i}}{2|\xi_{i}|}K_{+}^{(i)}+\frac{\xi_{i}^{*}}{2|\xi_{i}|}K_{-}^{(i)}\right),\label{transf6}
\end{equation}
where $i=a,b$ and $\left(D^{\dag}K_{\pm}D\right)^{\dag}=D^{\dag}K_{\mp}D$. Moreover, the bosonic operators $\hat{a}$ and $\hat{b}$ are transformed in terms of this displacement operator as
\begin{equation}
D^{\dag}(\xi_{a})\hat{a}D(\xi_{a})=\hat{a}\cosh(|\xi_{a}|)+\hat{a}^{\dag}\frac{\xi_{a}}{|\xi_{a}|}\sinh(|\xi_{a}|),\quad\quad D^{\dag}(\xi_{b})\hat{b}D(\xi_{b})=\hat{b}\cosh(|\xi_{b}|)+\hat{b}^{\dag}\frac{\xi_{b}}{|\xi_{b}|}\sinh(|\xi_{b}|).\label{JB}
\end{equation}

\section*{Appendix B. The similarity transformations of the operator $i\frac{\partial}{\partial t}$}

Now, we shall compute the similarity transformation of the operator $i\frac{\partial}{\partial t}$ in terms of the displacement operators $D(\chi)$ and $D(\xi)_{ab}$ of equations (\ref{Dsu2}) and (\ref{Dsu11xsu11}), respectively. Thus, we proceed to apply these transformations in the following order
\begin{equation}
D^{\dag}(\xi(t))_{ab}D^{\dag}(\chi(t))\left(i\frac{\partial}{\partial t}\right)D(\chi(t))D(\xi(t))_{ab}=D^{\dag}(\xi(t))_{ab}\left(i\frac{\partial}{\partial t}\right)'D(\xi(t))_{ab}=\left(i\frac{\partial}{\partial t}\right)''.
\end{equation}

As it is shown in Ref. \cite{Nos6}, from the first transformation $\left(i\frac{\partial}{\partial t}\right)'$ we obtain
\begin{equation}
\left(i\frac{\partial}{\partial t}\right)'=i\frac{\partial}{\partial t}+a_{0}J_{0}+a_{1}J_{+}+a_{1}^{*}J_{-},
\end{equation}
where the complex constants $a$'s are explicitly given as
\begin{equation}
a_{0}=\dot{\phi}\left(\cos(\theta)-1\right),\quad\quad a_{1}=-\frac{e^{-i\phi}}{2}\left(\dot{\phi}\sin(\theta)+i\dot{\theta}\right).
\end{equation}
With this result we obtain that the transformation $\left(i\frac{\partial}{\partial t}\right)''$ can be written as
\begin{equation}
D^{\dag}(\xi(t))_{ab}\left(i\frac{\partial}{\partial t}\right)'D(\xi(t))_{ab}=D^{\dag}(\xi(t))_{ab}\left(i\frac{\partial}{\partial t}\right)D(\xi(t))_{ab}+D^{\dag}(\xi(t))_{ab}\{a_{0}J_{0}+a_{1}J_{+}+a_{1}^{*}J_{-}\}D(\xi(t))_{ab}.\label{dt2}
\end{equation}
The first term of this expression results to be \cite{Nos6}
\begin{equation}
D^{\dag}(\xi(t))_{ab}\left(i\frac{\partial}{\partial t}\right)D(\xi(t))_{ab}=i\frac{\partial}{\partial t}+b_{0}K^{(ab)}_{0}+b_{1}J_{0}+b_{2}K^{(a)}_{+}+b^{*}_{2}K^{(a)}_{-}+b_{3}K^{(b)}_{+}+b^{*}_{3}K^{(b)}_{-},\label{DdtD}
\end{equation}
with the values of the complex constants $b$'s given by
\begin{equation}
b_{0}=\frac{1}{2}\left(\dot{\phi}_{a}(\cosh(\theta_{a})-1)+\dot{\phi}_{b}(\cosh(\theta_{b})-1)\right),
\quad\quad b_{1}=\frac{1}{2}\left(\dot{\phi}_{a}(\cosh(\theta_{a})-1)-\dot{\phi}_{b}(\cosh(\theta_{b})-1)\right),
\end{equation}
\begin{equation}
b_{2}=-\frac{e^{-i\phi_{a}}}{2}\left(\dot{\phi}_{a}\sinh(\theta_{a})+i\dot{\theta}_{a}\right),\quad\quad b_{3}=-\frac{e^{-i\phi_{b}}}{2}\left(\dot{\phi}_{b}\sinh(\theta_{b})+i\dot{\theta}_{b}\right).
\end{equation}

On the other hand, by using the expressions (\ref{JB}) we can calculate the second term of the expression (\ref{dt2}), which is given by
\begin{align}
D^{\dag}(\xi(t))_{ab}\{a_{0}J_{0}+a_{1}J_{+}+a_{1}^{*}J_{-}\}D(\xi(t))_{ab}&=c_{0}K^{(ab)}_{0}+c_{1}J_{0}+c_{2}J_{+}+c^{*}_{2}J_{-}+c_{3}K^{(ab)}_{+}\nonumber \\&+c^{*}_{3}K^{(ab)}_{-}+c_{4}K^{(a)}_{+}+c^{*}_{4}K^{(a)}_{-}+c_{5}K^{(b)}_{+}+c^{*}_{5}K^{(b)}_{-}.\label{DJD}
\end{align}
Here, the complex constants $c$'s are explicitly given by the expressions
\begin{equation}
c_{0}=a_{0}\left(\frac{\cosh(\theta_{a})-\cosh(\theta_{b})}{2}\right),\quad\quad c_{1}=a_{0}\left(\frac{\cosh(\theta_{a})+\cosh(\theta_{b})}{2}\right),
\end{equation}
\begin{equation}
c_{2}=a_{1}\cosh\left(\frac{\theta_{a}}{2}\right)\cosh\left(\frac{\theta_{b}}{2}\right)+a^{*}_{1}e^{-i(\phi_{a}-\phi_{b})}\sinh\left(\frac{\theta_{a}}{2}\right)\sinh\left(\frac{\theta_{b}}{2}\right),
\end{equation}
\begin{equation}
c_{3}=-\left[a_{1}e^{-i\phi_{b}}\cosh\left(\frac{\theta_{a}}{2}\right)\sinh\left(\frac{\theta_{b}}{2}\right)+a^{*}_{1}e^{-i\phi_{a}}\cosh\left(\frac{\theta_{b}}{2}\right)\sinh\left(\frac{\theta_{a}}{2}\right)\right],
\end{equation}
\begin{equation}
c_{4}=-\frac{\sinh(\theta_{a})}{2}e^{-i\phi_{a}},\quad\quad c_{5}=\frac{\sinh(\theta_{b})}{2}e^{-i\phi_{b}}.
\end{equation}
Therefore, from the results of the expressions (\ref{DdtD}) and (\ref{DJD}) we have that the equation (\ref{dt2}) can be written as
\begin{align}
\left(i\frac{\partial}{\partial t}\right)''&=i\frac{\partial}{\partial t}+ (b_{0}+c_{0})K_{0}^{(ab)}+(b_{1}+c_{1})J_{0}+c_{3}J_{+}+c_{3}^{*}J_{-}+c_{4}K_{+}^{(ab)}+c_{4}^{*}K_{-}^{(ab)}\nonumber\\&+(b_{2}+c_{5})K_{+}^{(a)}+(b_{2}^{*}+c_{5}^{*})K_{-}^{(a)}+(b_{3}+c_{6})K_{+}^{(b)}+(b_{3}^{*}+c_{6}^{*})K_{-}^{(b)}.
\end{align}

\end{document}